\documentstyle[epsfig]{aipproc}
\def\Journal#1#2#3#4{{\em #1}{\bf #2}, #3 (#4)}

\def\NPA{Nucl.~Phys. \bf{A}}

\def\PRD{Phys.~Rev. \bf{D}}

\def\PR{Phys.~Rev.~}

\begin{document}
\title{Relation between Scattering and 
Production Amplitude\\
{\large ---Case of Intermediate $\sigma$-Particle
   in $\pi\pi$-System---  }
}
\author{
Muneyuki Ishida$^*$, Shin Ishida$^{\dagger}$ and Taku Ishida$^{\ddagger}$ }
\address{
$^*$
Department of Physics, University of Tokyo, Tokyo 113,
$^{\dagger}$
Atomic Energy Research Institute, College of Science and Technology,
Nihon University, Tokyo 101 and 
$^{\dagger}$
KEK, Oho, Tsukuba, Ibaraki 305, JAPAN}

%\lefthead{LEFT head}
%\righthead{RIGHT head}
\maketitle

\begin{abstract}
The relation between scattering and production 
amplitudes are 
investigated, using a simple field theoretical model,
 from the general viewpoint of unitarity and 
the applicability of
final state interaction(FSI-) theorem.
%, especially
%in relation to the universality argument.
%As a result it is shown that 
The IA-method and VMW-method,
which are applied to our phenomenological 
analyses\cite{rf:had97a,rf:had97d}
suggesting the $\sigma$-existence,  are 
obtained as the physical state representations of
scattering and production amplitudes, respectively.
Moreover,
the VMW-method is shown to be an effective
method to obtain the resonance properties
from general production processes,
% generally with the 
%unknown strong-phases,
while the conventional analyses based on the 
``universality'' of $\pi\pi$-scattering amplitude 
are powerless for this purpose.

\end{abstract}

%%%%%%%%%%%%%%%%%%%%%%%%%%%%%%%%%%%%%%%%%%%%%%%%%%%%%%%%%%%%%%%%%%%%
%\noindent 
({\em General problem})\ \ \ \ 
In our phenomenological analyses suggesting the 
$\sigma$-existence 
the $\pi\pi$-scattering amplitude ${\cal T}$ and 
the production  amplitude ${\cal F}$ are parametrized 
by IA-method and VMW-method, respectively.
In treating  ${\cal T}$ and  ${\cal F}$ there are two general problems to be taken into account: 
\ \  The  ${\cal T}$ must satisfy
the unitarity,\ 
$
{\cal T}-{\cal T}^\dagger = 2i{\cal T}\rho{\cal T}^\dagger .
$\ \ 
and the 
 ${\cal F}$ must have the same 
phase\cite{rf:watson} as  ${\cal T}$:  \ 
$
{\cal T}\propto e^{i\delta} \rightarrow  
{\cal F}\propto e^{i\delta},
$
in case that the initial states have 
no strong phases. 
Moreover,
on the basis of the ``Universality"\cite{rf:pennington,rf:morg}
%of $\pi\pi$-scattering,
the more restrictive  relation than FSI-theorem 
between  ${\cal F}$ and  ${\cal T}$ is  required: 
%\begin{eqnarray}
$
{\cal F} = \alpha (s){\cal T}
$
%\label{eq:FaT}
%\end{eqnarray}
with a slowly varying real function $\alpha (s)$ of $s$.

We re-examine the relation
between ${\cal F}$ and ${\cal T}$ 
concretely, by using a simple model\cite{rf:aitchson,rf:chung}:
The pion $\pi$ and
the resonant particles such as $\sigma$(600) or $f_0(980)$  
are introduced equally as bare states, 
 denoted as $|\bar\alpha\rangle =|\bar \pi\rangle ,\ 
|\bar \sigma\rangle ,
\  |\bar f\rangle $, which 
are stable particles with zero widths.
By taking into account
 the residual strong interaction between these color-singlet states
(and a production channel ``$P$''),
 \begin{eqnarray}
{\cal L}^{\rm scatt}_{\rm int} &=&
\sum \bar g_\alpha\bar\alpha\pi\pi
+\bar g_{2\pi}(\pi )^4\ \ \ \  
({\cal L}^{\rm prod}_{\rm int}=\sum\bar\xi_\alpha\bar\alpha ``P"
+\bar\xi_{2\pi}\pi\pi  ``P") ,
\label{eq:Lint}
\end{eqnarray}
the bare states change into the physical states
acquiring finite widths. 
In the following we consider only the 
 repetition of the  $\pi\pi$-loop effects.

%\noindent 
({\em 3-different ways of 
 description of scattering amplitudes})\ \ \ \ 
There are three different ways of description 
of scattering amplitudes, 
corresponding to the three sorts of basic 
states for describing the resonant particles:
the bare states $|\bar\alpha\rangle$, the ${\cal K}$-matrix states 
$|\tilde\alpha\rangle$, and the physical states  
$|\alpha\rangle$ with a definite mass and lifetime. 

First we consider the two( $\bar\sigma ,\ \bar f$) 
resonance-dominative case.
The ${\cal T}$ is represented in terms of 
the $\pi\pi$-coupling constants $\bar g_{\bar\alpha}$
 and the propagator matrix $\bar\Delta$ as
\begin{eqnarray}
{\cal T}^{\rm Res} &=&
 \bar g_{\bar\alpha}\bar\Delta_{\bar\alpha\bar\beta}
\bar g_{\bar\beta};\ \ \ 
\bar\Delta_{\bar\alpha\bar\beta}^{-1}
 = (\bar M^2-s-i\bar G)_{\bar\alpha\bar\beta},\ \ 
\bar G_{\bar\alpha\bar\beta}
 = \bar g_{\bar\alpha}\rho\bar g_{\bar\beta}.
\label{eq:TinB}
\end{eqnarray}
This ${\cal T}^{\rm Res}$ is easily shown to satisfy 
the unitarity.

In the following we start from the ``${\cal K}$-matrix" states,
which are able to be identified
with the bare states
$|\bar \alpha\rangle (\equiv |\tilde \alpha\rangle )$,
and suppose  
$(\bar M^2-s)_{\bar\alpha\bar\beta}=
(\bar m^2_{\bar\alpha}-s)\delta_{\bar\alpha\bar\beta}$,
without loss of essential points.\footnote{
The bare states are related to the ``${\cal K}$-matrix''
states through the orthogonal transformation, which does not
 change
the reality of coupling constant.
The real part of the mass correction generally do not
have sharp s-dependence. Thus,
 the coupling constant in the ``${\cal K}$-matrix'' representation 
remains almost $s$-independent except for the threshold region.
 } 
The ${\cal T}^{\rm Res}$  can be expressed in the form
representing concretely 
the repetition of the $\pi\pi$-loop, as 
\begin{eqnarray}
{\cal T}^{\rm Res} 
  &=& 
{\cal K}^{\rm Res}/(1-i\rho{\cal K}^{\rm Res});\ \ \ \ 
{\cal K}^{\rm Res}
=\bar g_{\bar\sigma}(\bar m_{\bar\sigma}^2-s)^{-1}\bar g_{\bar\sigma}
+\bar g_{\bar f}(\bar m_{\bar f}^2-s)^{-1}\bar g_{\bar f}.
\label{eq:Kresrep}
\end{eqnarray}
This is the same form as the conventional 
${\cal K}$-matrix in potential theory. 
From the viewpoint of the present field-theoretical model, 
this ``${\cal K}$-matrix" has a physical meaning
as the  propagators of bare particles
with infinitesimal imaginary widths,
$\bar m_{\bar\alpha}^2\rightarrow
\bar m_{\bar\alpha}^2-i\epsilon$,
while the original 
${\cal K}$-matrix in potential theory is purely real
and has no direct meaning.

%\noindent 
({\em Relation between ${\cal T}$ and ${\cal F}$})\ \ \ \ 
The production amplitude ${\cal F}$ is obtained,
by the substitution, $\bar g^2\rightarrow\bar g\bar\xi$
 in the numerator ${\cal K}^{\rm Res}$ in Eq.(\ref{eq:Kresrep})
($\bar\xi$ being  the production coupling-constant), as
\begin{eqnarray}
{\cal F}^{\rm Res} &=& {\cal P}^{\rm Res}/(1-i\rho{\cal K}^{\rm Res});\ \ \ \ 
{\cal P}^{\rm Res}=\bar\xi_{\bar\sigma}
(\bar m_{\bar\sigma}^2-s)^{-1}\bar g_{\bar\sigma}
+\bar\xi_{\bar f}
(\bar m_{\bar f}^2-s)^{-1}\bar g_{\bar f}.
\label{eq:Presrep}
\end{eqnarray}
The FSI-theorem is automatically satisfied
since both\footnote{ 
The criticism  on our present work, raised by 
 M.R.Pennington\cite{rf:pen}, that a spurious zero 
of ${\cal T}$ transmits to ${\cal F}$ unphysically, 
is due to his  misunderstanding the relation between 
 Eqs.(\ref{eq:Kresrep}) and (\ref{eq:Presrep}).
The positions of zero in ${\cal T}$ and ${\cal F}$, 
which are determined through ${\cal K}^{\rm Res}=0$ and 
${\cal P}^{\rm Res}=0$, are  $s=s_0^{\cal T}=
(\bar g_{\bar\sigma}^2\bar m_{\bar f}^2+
\bar g_{\bar f}\bar m_{\bar \sigma}^2)
/(\bar g_{\bar\sigma}^2+\bar g_{\bar f}^2)$
 and   
$s=s_0^{\cal F}=
(\bar g_{\bar\sigma}\bar\xi_{\bar\sigma}\bar m_{\bar f}^2+
\bar g_{\bar f}\bar\xi_{\bar f}\bar m_{\bar\sigma}^2)
/(\bar g_{\bar\sigma}\bar\xi_{\bar\sigma}+
\bar g_{\bar f}\bar\xi_{\bar f})$, respectively.
%  
%, if they exist, have different positions,
The $s_0^{\cal F}$ are 
dependent on  the production couplings, $\bar\xi_{\bar\sigma}$
 and  $\bar\xi_{\bar f}$, in the respective processes,
% through ${\cal P}^{\rm Res}=0$,
%while they do not appear in the case
% $\bar g_{\bar\sigma}\bar\xi_{\bar\sigma}
%=-\bar g_{\bar f}\bar\xi_{\bar f}$.  
%They
and generally different from $s_0^{\cal T}$  
except for the special case $\bar\xi_{\bar\sigma}/
\bar g_{\bar\sigma}=\bar\xi_{\bar f}/\bar g_{\bar f}$.
%, 
%implicitly assumed\cite{rf:had97} by him.
 }
 ${\cal K}^{\rm Res}$ and ${\cal P}^{\rm Res}$
can be treated as real and
the phases of ${\cal T}^{\rm Res}$ and ${\cal F}^{\rm Res}$
come from the common factor $(1-i\rho{\cal K}^{\rm Res})^{-1}$.

By diagonalizing $\bar G$ in Eq.(\ref{eq:TinB}) 
further by a complex orthogonal matrix $u_{\bar\alpha\alpha}$ 
 we obtain 
the ${\cal T}^{\rm Res}$ in the physical state representation 
 given by
\begin{eqnarray}
{\cal T}^{\rm Res} &=& F_{\alpha}\Delta_{\alpha\beta}F_{\beta}
=F_\sigma (\lambda_\sigma-s)^{-1} F_\sigma  
+F_f(\lambda_f-s)^{-1}F_f;\ \ \ 
\label{eq:TinP}
\end{eqnarray}
where the $\lambda_\alpha$ is the physical squared mass of the
$\alpha$-state, and  the 
$F_{\alpha}$ 
is the coupling constant in physical state representation,
 which is generally complex.
By using 
 the real physical coupling $g_\alpha$ defined by
$g_\alpha^2\equiv -{\rm Im}\ \lambda_\alpha /\rho$,
the ${\cal T}^{\rm Res}$ is rewritten into the following form:
\begin{eqnarray}
{\cal T}^{\rm Res} &=& \frac{g_\sigma^2}{\lambda_\sigma-s}  
+\frac{g_f^2}{\lambda_f-s}+2i\rho 
\frac{g_\sigma^2}{\lambda_\sigma-s}\frac{g_f^2}{\lambda_f-s},
 \label{eq:IAderive}
\end{eqnarray}
where the $\lambda_\alpha$ and  $g_\alpha$ are 
represented by $\bar m_{\bar\alpha}$,  $\bar g_{\bar\alpha}$,
and the $\pi\pi$-state density
 $\rho (=\sqrt{1-4m_\pi^2/s}/16\pi )$, and accordingly are  
also almost $s$-independent except for 
the threshold region. Thus Eq.(\ref{eq:IAderive}) 
 is just the form of
scattering amplitude, applied in IA-method.
 
Similarly the ${\cal F}^{\rm Res}$ 
in the physical state representation
is given by
\begin{eqnarray}
{\cal F}^{\rm Res} &=& 
\frac{r_\sigma e^{i\theta_\sigma}}{\lambda_\sigma-s}  
+\frac{r_f e^{i\theta_f}}{\lambda_f-s},
 \label{eq:VMWderive}
\end{eqnarray}
where the $r_\sigma e^{i\theta_\sigma}\equiv
\Sigma_\sigma F_\sigma$
and 
$r_f e^{i\theta_f}\equiv
\Sigma_f F_f$(,$\Sigma_\alpha$ being
the complex physical production coupling defined by
$\Sigma_\alpha\equiv \bar g_{\bar\beta}u_{\bar\beta\alpha}$).
The $r_\sigma ,r_f, \theta_\sigma$ and $\theta_f$
are given by
\begin{eqnarray}
r_\alpha e^{i\theta_\alpha} &=&
% /(\lambda_\beta -\lambda_\alpha )
[\bar r_{\bar\alpha }
(\bar m_{\bar \beta}^2-\lambda_\alpha )
+\bar r_{\bar \beta}
(\bar m_{\bar\alpha}^2-\lambda_\alpha )]
 /(\lambda_\beta -\lambda_\alpha );\ \ 
\bar r_{\bar\alpha}\equiv \bar g_{\bar\alpha}\bar\xi_{\bar\alpha},
\label{eq:con1}
\end{eqnarray}
where $(\alpha ,\beta )=(\sigma ,f)\ {\rm or}\ (f,\sigma)$.
As can be seen by Eq.(\ref{eq:con1})
the $r_\alpha$ and $\theta_\alpha$ are almost 
$s$-independent except for the threshold region.
Thus,
the Eq.(\ref{eq:VMWderive}) is just the same formula 
 applied in VMW-method\cite{rf:had97d}.

In the VMW-method essentially the 
three new parameters,
$r_\sigma ,\ r_f$ and the relative phase 
$\theta (\equiv\theta_\sigma -\theta_f)$,
independent of the 
scattering process,
characterize the relevant production processes.
Presently they
are represented by the 
two production coupling constants, $\bar\xi_{\bar\sigma}$
and $\bar\xi_{\bar f}$ (or equivalently $\bar r_{\bar\sigma}$
and $\bar r_{\bar f}$).
Thus, among the three parameters in VMW-method there is
one constraint due to the FSI-theorem.

%\noindent 
({\em Background effect})\ \ \ \ 
Next we consider the effect of the non-resonant 
 background phase $\delta_{\rm BG}$.
Applying a general prescription in the  
 IA-method\cite{rf:had97a}, 
 the 
%${\cal T}$ the  ${\cal F}$ 
amplitudes are obtained 
in a similar manner to Eq.(\ref{eq:Presrep}), as
\begin{eqnarray}
{\cal T} = 
\frac{{\cal K}^{\rm Res}+{\cal K}^{\rm BG}}
{(1-i\rho{\cal K}^{\rm Res})(1-i\rho{\cal K}^{\rm BG})}
  & \rightarrow &
{\cal F} =
\frac{{\cal P}^{\rm Res}+{\cal P}^{\rm BG}}
{(1-i\rho{\cal K}^{\rm Res})(1-i\rho{\cal K}^{\rm BG})},
\label{eq:TBG}
\end{eqnarray}
where  ${\cal K}^{\rm Res}$ and
 ${\cal K}^{\rm BG}$ ( ${\cal P}^{\rm Res}$ and
 ${\cal P}^{\rm BG}$ ) is, respectively, 
the resonant and background ${\cal K}$-matrix
in scattering (production) process.
The  ${\cal K}^{\rm BG}$(${\cal P}^{\rm BG}$) 
is equal to the background coupling 
$\bar g_{2\pi}(s)(\bar\xi_{2\pi}(s)$).
This ${\cal F}$ automatically satisfies the FSI-theorem.
The ${\cal F}$ is rewritten into the same form\footnote{
Moreover, the ${\cal F}$ has the 
 overall phase factor 
$e^{i\delta_{\rm BG}}$
which plays a role only in the angular analysis
through the scalar-tensor interfering term.
This factor has a dull $s$-dependence
and its effect may be 
regarded as being
included in the phase parameters,
the $\theta_\alpha$, of VMW-method.
 }
as Eq.(\ref{eq:VMWderive}) of VMW-method,  
 except for the $s$-dependence of 
production couplings $r_\sigma (s),\theta_\sigma (s), r_f(s),$
 and $\theta_f(s)$, which are obtained by 
substituting to the 
$\bar r_{\bar\alpha}(\bar\alpha =\bar\sigma ,\bar f)$
 in Eq.(\ref{eq:con1}) the $\bar r_{\bar\alpha}(s)$:
\begin{eqnarray}
\bar r_{\bar\sigma}(s) &=&
\bar r_{\bar\sigma}\ cos \delta_{\rm BG}+f_{\rm BG}(s)
(\bar m_{\bar\sigma}^2-s),\ \ 
\bar r_{\bar f}(s) =
\bar r_{\bar f}\ cos \delta_{\rm BG},
\label{eq:rs}
\end{eqnarray}
where the $f_{\rm BG}(s)$ is defined by 
$f_{\rm BG}(s)e^{i\delta_{\rm BG}}\equiv {\cal F}^{\rm BG}
(={\cal P}^{\rm BG}/(1-i\rho {\cal K}^{\rm BG}))$.
In the case  where 
the production coupling $\bar\xi_{2\pi}$ is so small
as $\bar r_{\bar\sigma}\gg f_{\rm BG}\bar m_{\bar\sigma}^2$,
the
$\bar r_{\bar\sigma}(s)$ and  $\bar r_{\bar f}(s)$ 
have dull $s$-dependence, and
are approximated with constants,
$\bar r_{\bar\sigma}(\bar m_{\bar\sigma}^2)$ and  
$\bar r_{\bar f}(\bar m_{\bar f}^2)$,
respectively, 
and the VMW-method
with  constrained phase parameters
is reproduced effectively 
in this case with a non-resonant $\delta_{\rm BG}$.  

%%%%%%%%%%%%%%%%%%%%%%%%%%%%%%%%%%%%%%%%%%%%%%%

%\noindent 
({\em Applicability of FSI-theorem})\ \ \ \ 
Here it should be noted that 
the FSI-theorem is only applicable
to the case of the initial state having no strong phase
 $\bar\theta^{\rm str.}$.
However, the initial $\bar\theta^{\rm str.}$ 
 exists generally in all processes 
under the effect of strong interactions.
%\footnote{
%On the other hand, in weak decays, such as $K\rightarrow 2\pi$ and
%$K_{l4}$ decays, the FSI-condition is exactly satisfied
%and the analysis by VMW-method with
%free parameters is not applicable
%to these processes.}
For example,
  the $pp$-central collision
%, which  is
%usually considered to be mainly due to 
% the double pomeron
%exchange process with no
%strong phase, 
is largely affected by the
$\Delta$-production.  
%causing the initial strong phase. 
 The $J/\Psi\rightarrow\omega\pi\pi$-decay process 
is also affected by the $b_1$-resonance effect.
%which supply with the initial strong phase.
In these cases we have few knowledge on the initial 
phases, and we are forced to treat the parameters 
in VMW-method as being effectively free.\footnote{
The effect of this unknown strong phase\cite{rf:chung} is 
able to be introduced in the VMW-method by
substitution of 
$
\bar r_{\bar\alpha} \rightarrow  \bar r_{\bar\alpha}
e^{i\bar\theta_{\bar\alpha}^{\rm str.}}.
$
}
The analyses presented by K.Takamatsu\cite{rf:had97d} 
 were done from this standpoint.

%%%%%%%%%%%%%%%%%%%%%%%%%%%%%%%%%%%%%%%%%%%%%%%

%\noindent 
({\em Physical meaning of ``Universality''})\ \ \ \ 
In the ``Universality'' argument 
the masses and widths of resonances are determined only from the 
$\pi\pi$-scattering. In actual analyses 
the $\alpha (s)$ is  arbitrarily chosen as,   
$\alpha (s) = \sum_{n=0} \alpha_ns^n, $
 and the analyses of respective production processes 
become nothing but the determination of  
the $\alpha_n$, which has no direct physical meaning. 

On the other hand 
in the VMW-method,
the difference between the spectra of
${\cal F}$ and ${\cal T}$ is explained
intuitively by taking the relations such as
$
\bar\xi_{\bar\sigma}/\bar g_{\bar\sigma}  \gg 
\bar\xi_{2\pi}/\bar g_{2\pi},
$
that is, the background effects are comparatively weaker in 
the production processes than in the scattering process.
Thus in this case the large low-energy peak structure 
in $|{\cal F}|^2$ 
shows directly the $\sigma$-existence.
In this situation
the properties of $\sigma$
can be obtained more exactly 
in the production processes
than in the scattering process,
and the VMW-method is  
effective for this purpose.
(See, more detail in the contribution\cite{rf:had97}.) 

%%%%%%%%%%%%%%%%%%%%%%%%%%%%%%%

%%%%%%%%%%%%%%%%%%%%%%%%%%%%%%%%%%%%%%%%%%%%%%%%%%%%%%%%%%%%%%%%%%%%

\vspace*{-0.4cm}
%%%%%%%%%%%%%%%%%%%%%%%%%%%%%%%%%%%%%%%%%%%%%%%%%%%%%%%%

\end{document}